\documentstyle[prl,aps,epsfig,multicol]{revtex}

\newcommand{\mc}{\multicolumn}  
\newcommand{\lsim}{\mathrel{\mathop{\kern 0pt \rlap
  {\raise.2ex\hbox{$<$}}}
  \lower.9ex\hbox{\kern-.190em $\sim$}}}
\newcommand{\gsim}{\mathrel{\mathop{\kern 0pt \rlap
  {\raise.2ex\hbox{$>$}}}
  \lower.9ex\hbox{\kern-.190em $\sim$}}}

\title{Nucleon--Nucleon Coincidence Spectra
in the Non--Mesonic Weak Decay of $\Lambda$--Hypernuclei and
the $\Gamma_n/\Gamma_p$ Puzzle}

\author{G. Garbarino, A. Parre\~{n}o and A. Ramos}

\address{Departament d'Estructura i Constituents de la Mat\`{e}ria,
Universitat de Barcelona, E--08028 Barcelona, Spain} \date{\today}

\begin{document}
\draft
\maketitle

\begin{abstract}

The main open problem in the physics of $\Lambda$--hypernuclei is the
lack of a sound theoretical interpretation of the large experimental
values for the ratio $\Gamma_n/\Gamma_p\equiv \Gamma(\Lambda n\to nn)/
\Gamma(\Lambda p\to np)$. To approach the problem, we have incorporated a 
$\Lambda N\to nN$ one--meson--exchange model in an intranuclear cascade code 
for a finite nucleus calculation of the nucleon--nucleon 
(angular and energy) coincidence distributions in 
the non--mesonic weak decay of $^5_\Lambda {\rm He}$ and 
$^{12}_\Lambda {\rm C}$ hypernuclei. 
The two--nucleon induced decay mechanism, $\Lambda np\to nnp$, has been taken 
into account by means of the polarization propagator formalism in 
local density approximation. With respect to single nucleon spectra studies, 
the treatment of two--nucleon correlations permits a cleaner and more direct 
determination, from data, of $\Gamma_n/\Gamma_p$. 
%We find that both the nucleon final state interactions and the two--nucleon induced decay
%channel play a crucial role in the determination of the distributions.
Our results agree with the preliminary coincidence data of KEK--E462
for $^5_\Lambda {\rm He}$. This allows us to conclude that
$\Gamma_n/\Gamma_p$ for $^5_\Lambda {\rm He}$ should be close
to $0.46$, the value predicted by our one--meson--exchange model
for this hypernucleus. Unfortunately, to disentangle one-- and two--body induced 
decay channels from experimental non--mesonic decay events, three--nucleon 
coincidences should be measured. 
%The simplistic picture that the back--to--back kinematics is able to 
%select one--nucleon induced processes is in fact far from being realistic.

\end{abstract}

% 21.80.+a   Hypernuclei 
% 13.30.Eg   Hadronic decays
% 13.75.Ev   Hyperon-nucleon interactions 
% 25.40.-h   Nucleon-induced reactions

\begin{multicols}{2}

%%%%%%%%%%%%%%%%%%%%%%%%%%%%%%%%%%%%%%%%%%%%%%%%%%%%%%%%%%%%%%%%%%%%%%
% INTRODUCTION
%%%%%%%%%%%%%%%%%%%%%%%%%
Since many years, a sound theoretical explanation of the large
experimental values of the ratio, $\Gamma_n/\Gamma_p$, 
between the neutron-- and proton--induced non--mesonic
decay widths, $\Gamma(\Lambda n\to nn)$ and $\Gamma(\Lambda p\to np)$,
of $\Lambda$--hypernuclei is missing \cite{Al02,Ra98}.
The calculations underestimate the central data for
all considered hypernuclei,  
%\[
%\left[\frac{{\Gamma}_n}{{\Gamma}_p}\right]^{\rm Th}\ll
%\left[\frac{{\Gamma}_n}{{\Gamma}_p}\right]^{\rm Exp} ,
%\hspace{2mm}
%0.5\lsim \left[\frac{{\Gamma}_n}{{\Gamma}_p}\right]^{\rm Exp}\lsim 2
%\]
although the large experimental error bars do not allow one to reach any
definite conclusion.  Moreover, in the experiments performed
until now it has not been possible to 
distinguish between nucleons produced by the one--body induced and the
(non--negligible) two--body induced decay mechanism,
$\Lambda NN\to nNN$. Because of its strong tensor 
%and weak central and parity--violating 
component, the one--pion--exchange (OPE) model with the ${\Delta}I=1/2$ isospin
rule supplies very small ratios, typically in the interval $0.05\div 0.20$. 
On the contrary, the OPE description can reproduce the total 
non--mesonic decay rates observed for light and medium hypernuclei.
Other interaction mechanisms beyond OPE might then be responsible for the 
overestimation of $\Gamma_p$ and the underestimation of $\Gamma_n$.
Those which have been studied extensively in the literature are the 
following ones: 1) the inclusion in the ${\Lambda}N\rightarrow nN$
transition potential of mesons heavier than the pion (also including the exchange
of correlated or uncorrelated two--pions)
\cite{Du96,Pa97,Os01,Pa01,It98}; 2) the inclusion of interaction terms that
explicitly violate the ${\Delta}I=1/2$ rule
\cite{Al02,Pa98,Al99b}; 3) the inclusion of the two--body induced decay mechanism
\cite{Ra95,Al00,Al99a} and 4) the description of the
short range $\Lambda N\to nN$ transition in terms of quark degrees of freedom
\cite{Ok99,Sa02}, which automatically introduces $\Delta I=3/2$ contributions.    

Recent progress has been made on the subject. \\
(1) On the one hand, a few calculations \cite{Os01,Pa01,It98,Ok99}
with $\Lambda N \rightarrow nN$ transition potentials including
heavy--meson--exchange and/or direct quark contributions
obtained ratios more in agreement with data, without providing, nevertheless, a
satisfactory explanation of the puzzle. In particular, these calculations
found a reduction of the proton--induced decay width due to the 
opposite sign of the tensor component of $K$--exchange with respect to the one
for $\pi$--exchange. Moreover, the parity violating
$\Lambda N(^3S_1)\to nN(^3P_1)$ transition, which contributes to both the $n$--
and $p$--induced processes, is considerably enhanced by $K$--exchange
and direct quark mechanisms and tends to increase $\Gamma_n/\Gamma_p$ \cite{Pa01,Ok99}.\\
(2) On the other hand, an error in the
computer program employed in Ref.~\cite{Ra97} to evaluate the single nucleon 
energy spectra from non--mesonic decay has been detected \cite{Ra02}. It consisted 
in the underestimation, by a factor ten, of the nucleon--nucleon collision
probabilities. The correction of such an error leads to quite different spectral 
shapes and allows one to extract smaller values of $\Gamma_n/\Gamma_p$ 
(which is a free parameter in the polarization propagator model of 
Refs.~\cite{Ra97,Ra02}) when a comparison
with old experimental data for $^{12}_\Lambda$C \cite{Mo74} is done.

%%%%%%%%%%%%%%%%%%%%%%%%%%%
% SCOPE OF THIS PAPER
%%%%%%%%%%%%%%%%%%%%%%%%%%%   
In the light of these recent developments and of new experiments
\cite{Ou00a,FI01,Gi01}, it is important to develop different theoretical
approaches and strategies for the determination of the $\Gamma_n/\Gamma_p$ 
ratio. In this Letter we present a finite nucleus calculation
of the nucleon--nucleon coincidence distributions in the
non--mesonic weak decay of $^5_\Lambda {\rm He}$ and $^{12}_\Lambda$C
hypernuclei. The work is motivated by the fact that, unlike the single
nucleon spectra, correlation observables permit a cleaner and more direct
extraction of $\Gamma_n/\Gamma_p$ from data \cite{Al02}.
An experiment performed very recently at KEK \cite{Ou00a} has actually
measured the angular and energy correlations that we discuss in this paper.
Some preliminary results of the experiment can already be
compared with our calculations.

The one--meson--exchange (OME) 
weak transition potential we employ to describe the one--nucleon stimulated
decays contains the exchange of $\rho$, $K$, $K^*$, $\omega$ and 
$\eta$ mesons in addition to the pion \cite{Pa01}. The final state interactions acting
between the two primary nucleons are taken into account by using a scattering $NN$ 
wave function from the Lippmann--Schwinger ($T$--matrix) equation obtained with 
the NSC97f potential \cite{nsc}. The OME decay rates predicted by this model are the
following ones \cite{Pa01}: $\Gamma_1\equiv \Gamma_n+\Gamma_p=0.32$,
$\Gamma_n/\Gamma_p=0.46$ for $^5_\Lambda$He and $\Gamma_1=0.55$, 
$\Gamma_n/\Gamma_p=0.34$ for $^{12}_\Lambda$C.

The distributions of primary nucleons emitted in 
two--nucleon induced decays and the ratio $\Gamma_2/\Gamma_1$ has been 
determined by the polarization propagator method
in local density approximation of Ref.~\cite{Al00}:
$\Gamma_2/\Gamma_1=0.20$ for $^5_\Lambda$He and 
$\Gamma_2/\Gamma_1=0.25$ for $^{12}_\Lambda$C \cite{prob}.

In their way out of the nucleus, the primary nucleons, 
due to collisions with other nucleons, continuously change energy, 
direction and charge. As a consequence, secondary nucleons are 
also emitted. We simulate the nucleon propagation inside the
residual nucleus with the Monte Carlo code of Ref.~\cite{Ra02}.

\begin{figure}
\begin{center}
\mbox{\epsfig{file=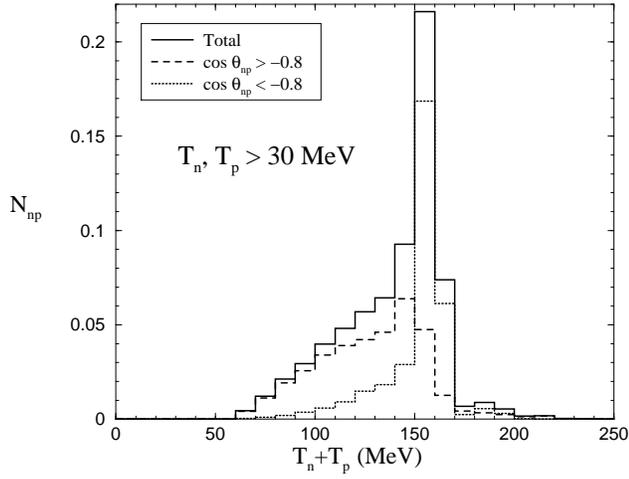,width=.45\textwidth}}
%\mbox{\epsfig{file=1.eps}}
\vskip 2mm
\caption{Kinetic energy correlations of $np$ pairs emitted {\it per
non--mesonic decay} of $^5_\Lambda$He. See text for details.}
\label{c1}
\end{center}
\end{figure}

\begin{figure}
\begin{center}
\mbox{\epsfig{file=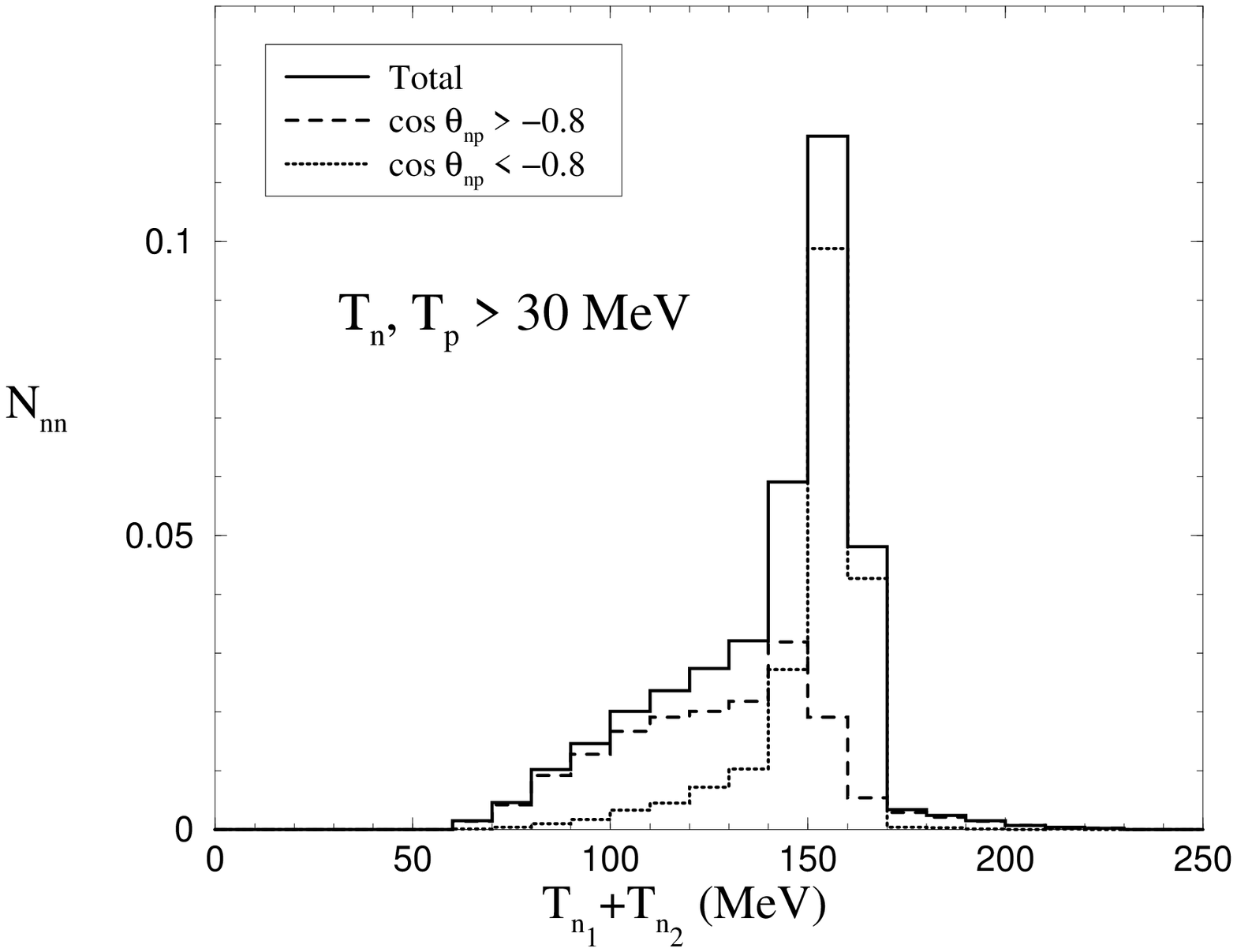,width=.45\textwidth}}
%\mbox{\epsfig{file=2.eps}}
\vskip 2mm
\caption{Kinetic energy correlations of $nn$ pairs emitted {\it per
non--mesonic decay} of $^5_\Lambda$He. See text for details.}
\label{c2}
\end{center}
\end{figure}

%%%%%%%%%%%%%%%%%
%   RESULTS
%%%%%%%%%%%%%%%%% 
In Fig.~\ref{c1} we show the kinetic energy correlation of $np$ 
coincidence pairs emitted in the non--mesonic
decay of $^5_\Lambda$He. The spectra
are normalized {\it per non--mesonic weak decay}.
To facilitate a comparison with experiments, 
whose kinetic energy threshold for proton (neutron) detection
is typically of about $30$ ($10$) MeV, and to avoid a possible 
non--realistic behaviour of the intranuclear cascade simulation 
for low nucleon energies, in all the figures of the paper
we required $T_n$, $T_p\geq 30$ MeV.
A narrow peak is predicted close 
%in the $150\div 160$ bin, close.. at $T_n+T_p\simeq 155$ MeV, close
to the $Q$--value (153 MeV) expected for the proton--induced three--body process
$^5_\Lambda{\rm He}\to ^3{\rm H}+n+p$: it 
is mainly originated by the back--to--back kinematics (cos$\, \theta_{np}<-0.8$).
A broad peak, predominantly due to $\Lambda p \to np$ or 
$\Lambda n \to nn$ weak transitions
followed by the emission of secondary (less energetic) nucleons,  
has been found around $140$ MeV for cos$\, \theta_{np}>-0.8$.
The kinetic energy correlation for $^5_\Lambda$He $nn$ pairs (Fig.~\ref{c2})
shows essentially the same structure of the $np$ distribution just discussed.

\begin{figure}
\begin{center}
\mbox{\epsfig{file=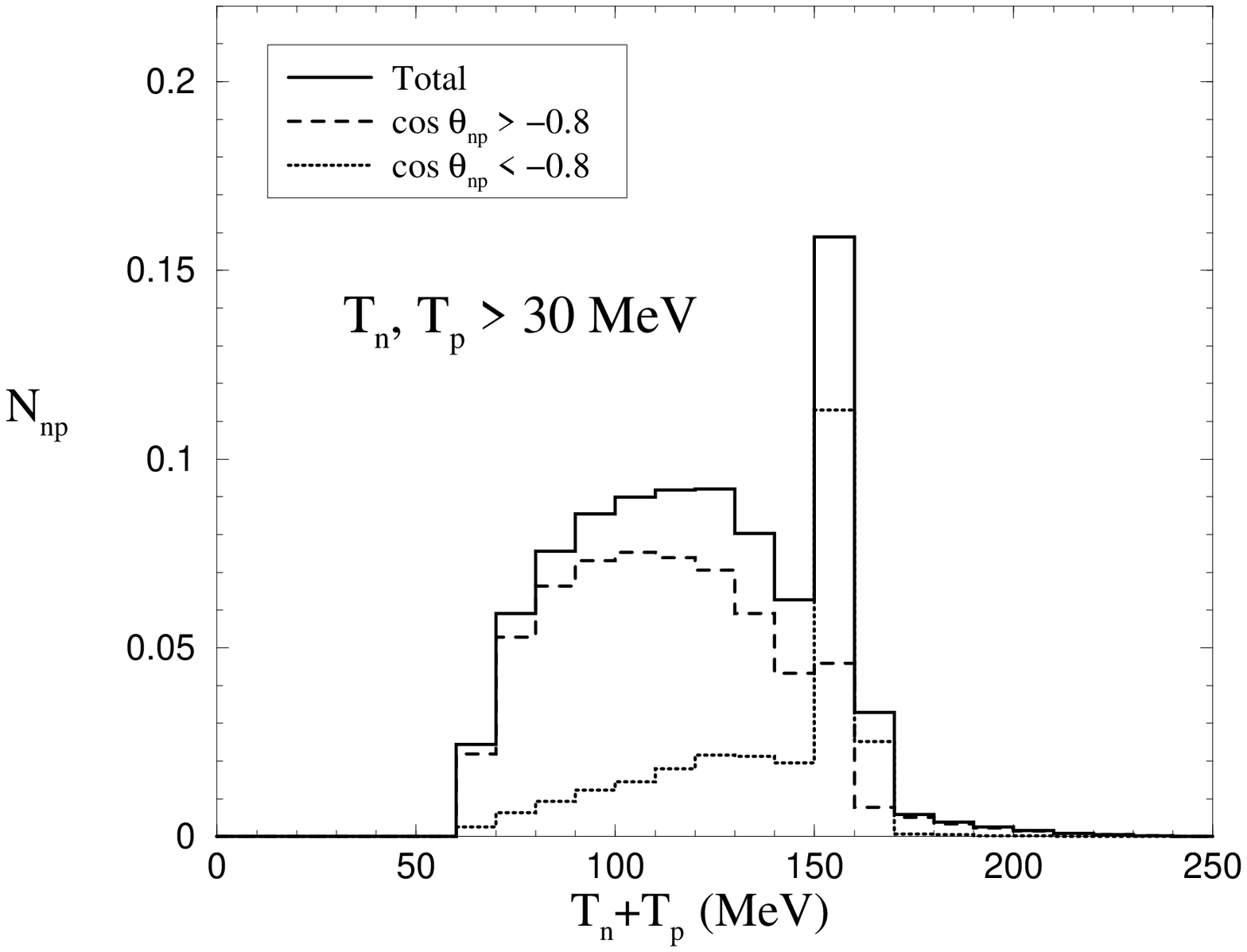,width=.45\textwidth}}
%\mbox{\epsfig{file=4.eps}}
\vskip 2mm
\caption{Kinetic energy correlations of $np$ pairs emitted 
{\it per non--mesonic decay} of $^{12}_\Lambda$C. See text for details.}
\label{c4}
\end{center}
\end{figure}

In Fig.~\ref{c4}, which corresponds to the energy correlation 
of $^{12}_\Lambda\rm{C}$ $np$ pairs, 
the narrow peak appearing at $T_n+T_p\simeq 155$ MeV again gets the dominant contribution
from back--to--back coincidences. The relevance of the nucleon final state 
interactions (FSI) in $^{12}_\Lambda$C relative to $^5_\Lambda$He can be seen
from the second, broader peak appearing in
the region around 110 MeV for $^{12}_\Lambda\rm{C}$ and 140 MeV for $^5_\Lambda{\rm He}$. 
This peak is in fact more pronounced for the heavier
hypernucleus. Another consequence of the different FSI effects in 
$^5_\Lambda$He and $^{12}_\Lambda$C is the different magnitude of the tail of the
back--to--back distribution at low energies.
%one corresponding to the
%$^{12}_\Lambda\rm{C} \to n+p+ ^{10}_\Lambda\rm{B}$ process 
%reduced by the significant FSI effects.

\begin{figure}
\begin{center}
\mbox{\epsfig{file=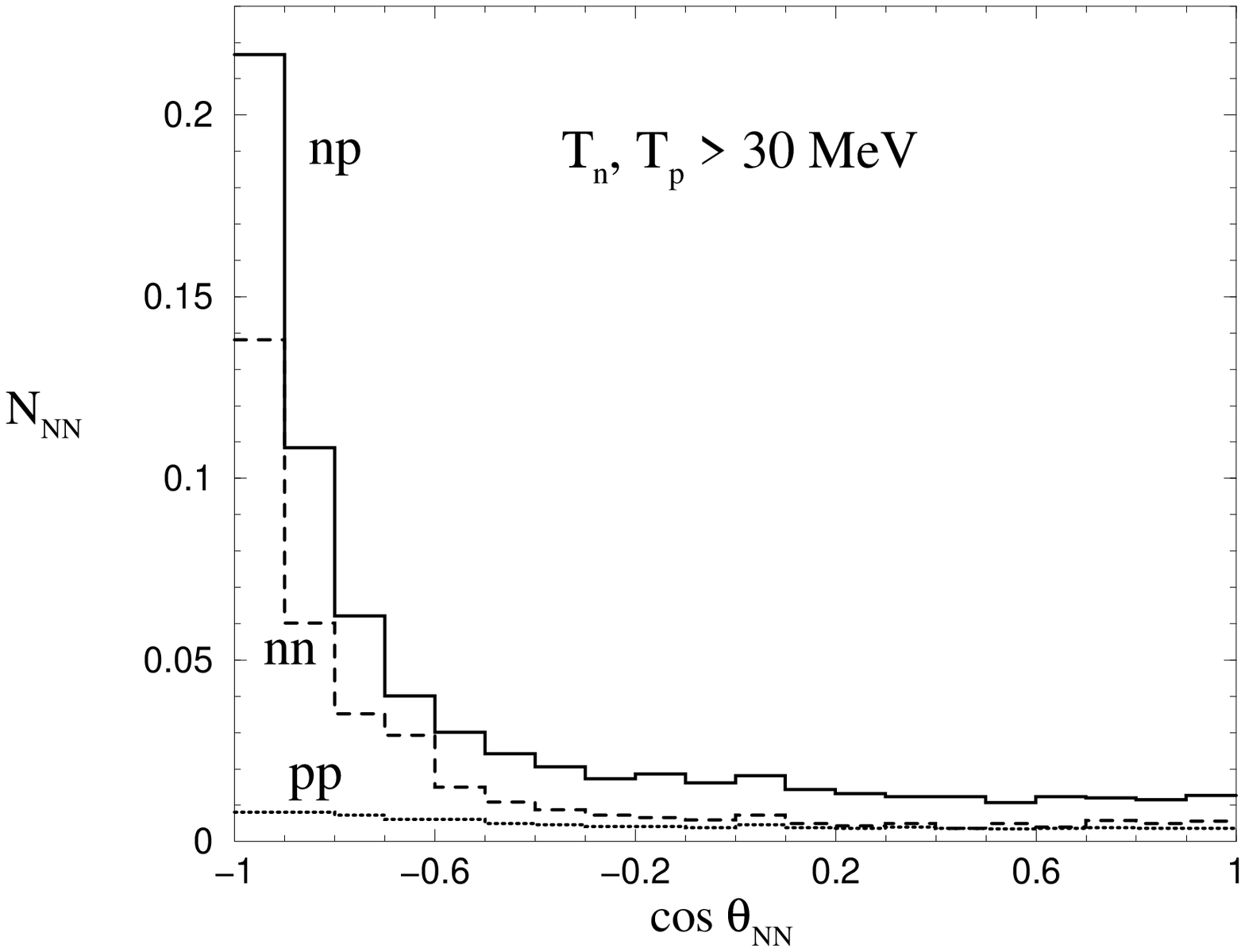,width=.45\textwidth}}
%\mbox{\epsfig{file=3.eps}}
\vskip 2mm
\caption{Opening angle correlations of $nn$, $np$ and $pp$ pairs
emitted {\it per non--mesonic decay} of $^5_\Lambda$He. See text for details.}
\label{c3}
\end{center}
\end{figure}

\begin{figure}
\begin{center}
\mbox{\epsfig{file=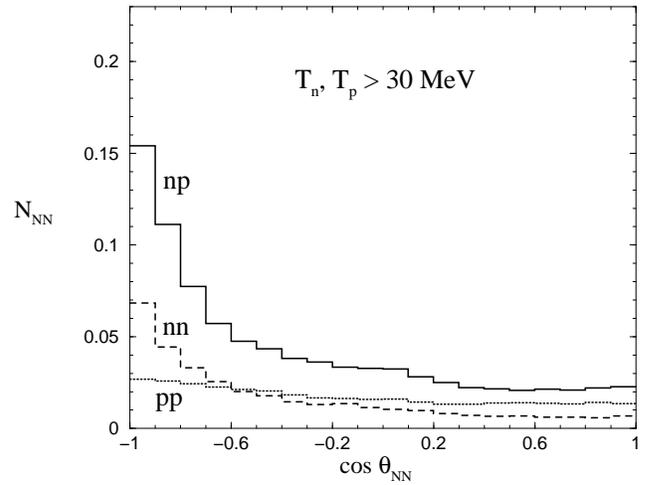,width=.45\textwidth}}
%\mbox{\epsfig{file=5.eps}}
\vskip 2mm
\caption{Opening angle correlations of $nn$, $np$ and $pp$ pairs
emitted {\it per non--mesonic decay} of $^{12}_\Lambda$C. See text for details.}
\label{c5}
\end{center}
\end{figure}

Figs.~\ref{c3} and \ref{c5} show the opening angle correlations of $nn, np$ and
$pp$ pairs emitted in the decay of $^5_\Lambda{\rm He}$ and 
$^{12}_\Lambda\rm{C}$, respectively. Comparing both figures for $nn$ and $np$
coincidences, one sees that the back--to--back peaks are more
pronounced for $^5_\Lambda{\rm He}$ 
(less sensitive to FSI than $^{12}_\Lambda\rm{C}$) than for 
$^{12}_\Lambda\rm{C}$, while the (almost uniform)
tail of these distributions (feeded by FSI) is more 
significant in $^{12}_\Lambda\rm{C}$ than in $^5_\Lambda{\rm He}$. 
Since at least one proton of any $pp$ coincidence is a secondary particle,
the $pp$ spectra are quite uniform: actually, due to the relevance of the back--to--back
kinematics in the weak decay, these distributions slowly decrease as
${\rm cos}\, \theta_{pp}$ increases. 
%This is due to the fact that at least
%a proton of the $pp$ pair is a secondary particle, then
%it is not far from being isotropically distributed with respect to the 
%other (primary or secondary) proton. 
Again as a consequence of FSI, the number of $pp$ pairs 
is considerably larger in $^{12}_\Lambda\rm{C}$ than in $^5_\Lambda{\rm He}$. 
%{\bf For $^{12}_\Lambda\rm{C}$ we have checked that about $91$\% ($98$\%) of the 
%total number of $pp$ pairs coming from the decay of $^{12}_\Lambda\rm{C}$ 
%have an energy per particle in the $0\div 30$ ($0\div 50$) MeV region.}

The ratio $\Gamma_n/\Gamma_p$ is defined as the ratio
between the number of primary weak decay $nn$ and $np$ pairs,
$N^{\rm wd}_{nn}$ and $N^{\rm wd}_{np}$. However, due to nucleon
FSI, one expects the inequality:
\begin{equation}
\label{ratio-nn}
\frac{\Gamma_n}{\Gamma_p}\equiv \frac{N^{\rm wd}_{nn}}{N^{\rm wd}_{np}}
\neq \frac{N_{nn}}{N_{np}}=f\left[\Delta \theta_{12}, \Delta(T_1+T_2)\right] 
\end{equation}
to be valid in a situation, such as the experimental one, in which particular intervals 
of variability of pair opening angle, $\Delta \theta_{12}$, and sum energy,
$\Delta(T_1+T_2)$, are employed in the determination of $N_{nn}$ and $N_{np}$.
Actually, as one can deduce from Figs.~\ref{c1}--\ref{c5},
not only $N_{nn}$ and $N_{np}$ but also the ratio $N_{nn}/N_{np}$ depends
on $\Delta \theta_{12}$ and $\Delta(T_1+T_2)$.
The numbers of nucleon pairs $N_{nn}$, $N_{np}$ and $N_{pp}$ discussed up to now
are related to the corresponding quantities for the one--nucleon ($N^{\rm 1B}_{NN}$)
and two--nucleon ($N^{\rm 2B}_{NN}$) induced processes
[the former (latter) being normalized per one--body (two--body) stimulated
non--mesonic weak decay] via the following equation:
\begin{equation}
\label{1-2}
N_{NN}=\frac{N^{\rm 1B}_{NN}\, \Gamma_1+N^{\rm 2B}_{NN}\, \Gamma_2}{\Gamma_1+\Gamma_2} .
\end{equation} 

Table \ref{sep-num} shows the dependence of $N_{nn}$, $N_{np}$ and $N_{nn}/N_{np}$
on $\Delta \theta_{12}$ and $\Delta(T_1+T_2)$ for $^{12}_\Lambda\rm{C}$. 
For comparison, the same quantities for the primary weak decay 
nucleons are listed as well. Without any restriction on $\theta_{NN}$ and the 
nucleon energies, one notes a great increase (by about one order of magnitude) 
of both the $nn$ and $np$ numbers when the effect of the FSI is taken into account. 
Again because of FSI, the use of an energy
threshold $T^{\rm th}_N$ of $30$ MeV supplies $N_{nn}$ and $N_{np}$ values considerably reduced
with respect to the ones obtained at $T^{\rm th}_N=0$ MeV. On the contrary, the ratio between
the number of $nn$ and $np$ pairs is much less sensitive to FSI effects.
\vskip -6mm
\vskip -6mm
\vskip -6mm    
\vskip -6mm  
\vskip -5mm
In Table \ref{ome-he} (\ref{ome-c}), the ratio $N_{nn}/N_{np}$
for $^5_\Lambda$He ($^{12}_\Lambda$C) is given for
different combinations of opening angle interval and nucleon energy threshold.
In parentheses we also report the predictions obtained when the
two--nucleon induced decay channel is neglected.
The results of the figures and tables presented in this work
are in a form that permits a direct
comparison with KEK--E462 data \cite{Ou00a}, which are now under analysis.
The first preliminary results of KEK--E462 are quoted in Table \ref{ome-he}.
We note that while the central values of the data are more in agreement with the
calculation that neglects the effect of the two--body induced decay mechanism,
the complete results are compatible with the upper limits of the same data. This 
could indicate an almost negligible effect of the $\Lambda np\to nnp$ process
and/or a $\Gamma_n/\Gamma_p$ ratio slightly lower than the one (0.46)
predicted by our OME model for $^5_\Lambda$He. 
To clarify this point, on the one hand one has
to wait for the final results of KEK--E462. On the other hand, our calculations
show that three--nucleon coincidences are required to disentangle the effects of
one-- and two--body stimulated decay channels from observed decay events. 
The simplistic picture that the back--to--back kinematics is able to
select one--nucleon induced processes is in fact far from being realistic. 

In conclusion, our OME weak interaction model
supplemented by FSI through an intranuclear cascade simulation
provides two--nucleon coincidence observables which reproduce
the preliminary KEK--E462 results for $^5_\Lambda$He. This allows us to 
conclude that $\Gamma_n/\Gamma_p$ for $^5_\Lambda {\rm He}$ should 
be close to $0.46$.
%In conclusion, our analysis of the preliminary KEK--E462 results for $^5_\Lambda$He
%supply a rather small $\Gamma_n/\Gamma_p$ ratio, around
%%In conclusion, in the light of our analysis, the preliminary results
%%of KEK--E462 are compatible with a rather small value of $\Gamma_n/\Gamma_p$,
%the value (0.46) predicted by the OME weak interaction model we employed. 
Although further (theoretical and experimental) confirmation is needed, 
in this paper we think we have proved how the study of nucleon
coincidence observables can offer a promising possibility
to solve the longstanding puzzle on the $\Gamma_n/\Gamma_p$ ratio.

In a forthcoming (long) paper \cite{new} we shall discuss
the nucleon correlation observables for the 
(one-- and two--nucleon stimulated) 
non--mesonic decay of $^5_\Lambda$He and $^{12}_\Lambda$C in a sistematic way. 
Single nucleon spectra will be further subject of this work.   
In addition, one should treat the case of
$^4_\Lambda$H, which is of extreme importance in order to test the validity 
of the $\Delta I=1/2$ isospin rule in the $\Lambda N\to nN$ weak transition
\cite{Al02,Al99b,Gi01}, another key point for the solution of the
$\Gamma_n/\Gamma_p$ puzzle.

This work is partly supported by EURIDICE HPRN--CT--2002--00311,
by the DGICYT BFM2002--01868, by the Generalitat de Catalunya SGR2001--64 and by
INFN. Discussions with H. Outa are acknowledged.

\begin{table}
\begin{center}
\caption{Results for $N_{nn}$, $N_{np}$ and $N_{nn}/N_{np}$ corresponding
to the non--mesonic decay of $^{12}_\Lambda$C.
A null (30 MeV for the numbers in parentheses) nucleon energy threshold and
two different opening angle regions are considered.}
\label{sep-num}
\begin{tabular}{c|c c}
\mc {1}{c|}{} &
\mc {1}{c}{cos$\,\theta_{NN}\leq -0.8$} &
\mc {1}{c}{all $\theta_{NN}$} \\ \hline
$N^{\rm wd}_{nn}$                    & $0.20$ ($0.19$) & $0.25$ ($0.24$)  \\ 
$N^{\rm wd}_{np}$                    & $0.57$ ($0.56$) & $0.75$ ($0.72$)  \\ 
$N^{\rm wd}_{nn}/N^{\rm wd}_{np}\equiv \Gamma_n/\Gamma_p$    & $0.34$ $(0.34)$ & $0.34$ $(0.34)$  \\ \hline
$N_{nn}$                             & $0.44$ $(0.11)$ & $3.15$ $(0.33)$  \\
$N_{np}$                             & $1.05$ $(0.26)$ & $8.40$ $(0.87)$  \\
$N_{nn}/N_{np}$                      & $0.42$ $(0.43)$ & $0.38$ $(0.39)$  \\
\end{tabular}
\end{center}
\end{table}
\vskip -6mm
\begin{table}
\begin{center}
\caption{Predictions of $N_{nn}/N_{np}$ for $^5_\Lambda$He corresponding to
different nucleon thresholds $T^{\rm th}_N$ and pair opening angles. 
The numbers in parentheses correspond to calculations with $\Gamma_2=0$
in Eq.~(\ref{1-2}). The (preliminary) data are from KEK--E462 \protect\cite{Ou00a}.}
\label{ome-he}
\begin{tabular}{c|c c c c}
\mc {1}{c|}{} &
\mc {1}{c}{} &
\mc {1}{c}{cos$\,\theta_{NN}$} &
\mc {1}{c}{} &
\mc {1}{c}{} \\
$T^{\rm th}_N$ (MeV) & $\leq -0.8$ & $\leq -0.6$ & $\leq -0.4$ & all \\ \hline
%$0$     & $0.57$ ($0.47$) & $0.53$ ($0.47$) & $0.50$ ($0.47$)  & $0.44$ ($0.43$) \\ \hline
$30$    & $0.61$ ($0.52$) & $0.61$ ($0.51$) & $0.60$ ($0.50$)  & $0.54$ ($0.45$) \\ 
                          & $0.52 \pm 0.11$ & $0.50 \pm 0.10$  & $0.51 \pm 0.10$ &    \\ \hline
$50$    & $0.63$ ($0.52$) & $0.61$ ($0.51$) & $0.60$ ($0.51$)  & $0.56$ ($0.46$) \\ 
\end{tabular}
\end{center}
\end{table}    
\vskip -6mm
\begin{table}
\begin{center}
\caption{Same as in Table \ref{ome-he} for $^{12}_\Lambda$C.}
\label{ome-c}
\begin{tabular}{c|c c c c}
\mc {1}{c|}{} &
\mc {1}{c}{} &
\mc {1}{c}{cos$\,\theta_{NN}$} &
\mc {1}{c}{} & 
\mc {1}{c}{} \\
$T^{\rm th}_N$ (MeV) & $\leq -0.8$ & $\leq -0.6$ & $\leq -0.4$ & all \\ \hline
%$0$   & $0.42$ ($0.38$) & $0.41$ ($0.38$) & $0.40$ ($0.38$) & $0.38$ ($0.36$) \\ \hline
$30$  & $0.43$ ($0.37$) & $0.43$ ($0.37$) & $0.43$ ($0.37$) & $0.39$ ($0.35$) \\ \hline
$50$  & $0.41$ ($0.35$) & $0.40$ ($0.35$) & $0.40$ ($0.35$) & $0.38$ ($0.34$) \\ 
\end{tabular}
\end{center}
\end{table} 

\end{multicols}

\begin{references}
\bibitem{prob} The values of $\Gamma_2/\Gamma_1$ used in this paper
have been obtained with the method of Ref.~\cite{Al00} after
correcting a small (conceptual) error in the implementation of data  
on the $P$--wave pion--nucleus optical potential. For 
$^5_\Lambda {\rm He}$ and $^{12}_\Lambda {\rm C}$, such a correction
slightly decreases the values of $\Gamma_1$ while increases $\Gamma_2$
by about 20\%.
\end{references}
\end{document}